\documentclass[preprint]{elsarticle}

\usepackage{lineno,hyperref}
\modulolinenumbers[5]
\journal{Journal of Magnetism and Magnetic Materials}

\biboptions{sort&compress}

\usepackage[usenames]{color}
\usepackage{verbatim}
\expandafter\let\csname equation*\endcsname\relax 
\expandafter\let\csname endequation*\endcsname\relax
\usepackage{amsmath,amsthm,amssymb,multicol}
\usepackage{graphics}
\usepackage{float}
\usepackage{hyperref}
\usepackage{subfigure}
\usepackage{anysize}
\marginsize{2.5cm}{2.5cm}{1cm}{1cm} 

\usepackage{soul}
\soulregister\cite7
\soulregister\ref7
\soulregister\pageref7

\usepackage{comands}

\makeatletter
\def\ps@pprintTitle{%
 \let\@oddhead\@empty
 \let\@evenhead\@empty
 \def\@oddfoot{}%
 \let\@evenfoot\@oddfoot}
\makeatother

\begin{document}

\begin{frontmatter}
\title{Spin excitations of half-doped bilayer manganites: intermediate phase}
\author{Ivon R. Buitrago$^{1,2}$, Cecilia I. Ventura$^{2,3}$, 
and Luis O. Manuel$^4$}
\address{$^1$ Instituto Balseiro, Univ. Nac. de Cuyo and CNEA, 8400-Bariloche, Argentina}
\address{$^2$ (CONICET) Centro At\'omico Bariloche-CNEA, Av. Bustillo 9500, 8400-Bariloche, Argentina}
\address{$^3$ Universidad Nacional de R\'{\i}o Negro, 8400-Bariloche, Argentina}
\address{$^4$ Instituto de F\'{\i}sica Rosario (CONICET-UNR), Rosario, Argentina.}

\ead{ivonnebuitrago@cab.cnea.gov.ar.com}

\date{\textrm{\today}}

\begin{abstract}
The ground state of half-doped manganites involves intricate spin, charge and orbital orderings, which are difficult 
to discern experimentally. In this work, we resort to the theoretical analysis of the spin fluctuation spectrum of the half-doped bilayer Pr(Ca$_{0.9}$Sr$_{0.1}$)$_2$ Mn$_2$O$_7$ in order to get an insight of its electronic ground state. By means of the linear spin wave approximation, we compute the magnon dispersion for a family of localized spin models, which can describe several phases proposed for the ground state of half-doped manganites, like the intermediate one proposed by Efremov \emph{et al.} [{\it Nat.Mats.} {\bf 3} 853, 2004] along with its particular cases corresponding to Goodenough's CE phase [{\it Phys. Rev.} 100, 564, 1955] and Zener polaron or dimer phases. We obtain an excellent agreement between theory and experiment when the ground state is assumed to be a generalized Goodenough's CE phase, with a Mn-charge disproportionation inside the experimentally expected range. As essential ingredients for our improved fit of the upper and lower magnon branches measured around the gap, we identified two next-nearest neighbour exchange interactions between the planar Mn zig-zag chains, one for each type of Mn ion present. In connection with this finding, we revisited the magnetic excitations of the laminar related compounds, focusing on the upper magnon branches. Here we prove that their measurement  would provide the key to identify unambiguously the ground state present in the layered half-doped manganites.\\
\end{abstract}

\end{frontmatter}


\section{Introduction}
\label{sec.Introduction}
The doped manganite oxides have the general formula (A$_{1-x}$B$_x$MnO$_3$)$_n$BO, where A and B are trivalent rare-earth-metal and divalent alkaline-earth-metal ions, respectively, and $n$ is the number of perovskite blocks separated by rock salt-structure non-magnetic BO layers. Thus, $n=1$ represents the layered structure \LCom{A}{1-x}{B}{1+x}, $n=2$ the bilayer A$_{2-2x}$B$_{1+2x}$Mn$_2$O$_7$, and $n\to\infty$ the three-dimensional \Com{A}{1-x}{B}{x}. Due to the strong correlations between charge, orbital and spin degrees of freedom, these manganese oxides exhibit rich phase diagrams, that include ferromagnetic metallic phases, antiferromagnetic insulating phases with charge and orbital order, among others which are of technological interest~\cite{Kajimoto,Dagotto,Hemberger,Larochelle}. At low doping ($x\sim0.3$), some of these compounds exhibit the interesting phenomenon of colossal magnetoresistance (CMR), which has been found in three dimensional manganites \Com{A}{1-x}{(Sr,Ca)}{x}~\cite{Tokura} and bilayer manganites such as La$_{2-2x}$Sr$_{1+2x}$Mn$_2$O$_7$, and only above high magnetic fields in layered La$_{1-x}$Sr$_{1+x}$MnO$_4$~\cite{Moritomo,Kawano}.

In his seminal work of 1955, Goodenough~\cite{Goodenough} predicted the magnetic, charge and orbital ordering for \Com{La}{1-x}{B}{x}, valid in the whole doping range: $ 0\leqslant x \leqslant 1$, based on his hypothesis of covalent/semicovalent bonds and the Zener double-exchange~\cite{Zener} picture. In particular, at half-doping ($x=0.5$) he proposed the antiferromagnetic CE phase. This phase is characterized by an arrangement of charges and spins, such that in the MnO$_2$ planes the \Mn{(3.5-\delta)} and \Mn{(3.5+\delta)} ions, where $\delta=0.5$ is a measure of the Mn-charge disproportionation (CD), form a checkerboard pattern and equal charges are stacked one above the other in the direction perpendicular to the planes. The spins of \Mn{3} ($S_1=2$) and \Mn{4} ($S_2=3/2$) ions form ferromagnetic zig-zag chains, which are antiferromagnetically coupled among them, in the planes and between them. The corner sites in the zig-zag chains are occupied by \Mn{4}, whereas the bridge sites are occupied by \Mn{3}. This charge order (CO) seemed to agree with several experimental results~\cite{Radaelli,Murakami,Di-Matteo}, which also suggested the simultaneous orbital ordering of the $e_g$ electrons of \Mn{3}. 

However, experimental measurements on \Com{Pr}{0.6}{Ca}{0.4}~\cite{Aladine,Grenier,Hill} and \Com{Y}{0.5}{Ca}{0.5}~\cite{Winkler} yielded no evidence of charge ordering, and so the proposal of Goodenough's CE ground state for these compounds was questioned. This gave rise to an alternative electronic state proposal, the Zener polaron dimer phase (ZP)~\cite{Aladine}, which has been studied by {\it ab initio} calculation~\cite{Bastardis}. In this phase an $e_g$ electron is delocalized between pairs of nearest neighbour Mn ions, due to the double-exchange mechanism, and therefore, the Mn ions form spin dimers and they have an intermediate valence \Mn{3.5} ($\delta=0$).

The experimental characterization of the CO was further complicated by the finding of non-zero charge disproportionations, lower than in Goodenough's CE phase, in several half-doped manganites~\cite{Hill,Winkler,Herrero2004,Subias,Herrero2012}. For example, \Mn{3.42} and \Mn{3.58} charge states (i.e. $\delta=0.08$) were found in \Com{Nd}{0.5}{Sr}{0.5}~\cite{Herrero2004}, while a charge imbalance of $\Delta n=0.14e^-$ was found between the two Mn sites in \Com{Bi}{0.5}{Sr}{0.5}~\cite{Subias} and $\Delta n=0.15e^-$ for layered \LCom{La}{0.5}{Sr}{1.5}~\cite{Herrero2012}. 

Given the growing debate on the two main scenarios proposed for the ground state of half-doped manganites, Ventura and Alascio~\cite{Ventura+Alascio} studied the spin dynamics which would result from a charge ordered and the ZP dimer phases, calculating the respective magnons using an effective model of localized spins. Differences between the magnon dispersion of these phases were found, which would make it possible to distinguish them through inelastic neutron scattering (INS) experiments~\cite{Ventura+Alascio}. Indeed, now INS data exist in half-doped manganites, by which it was attempted to shed light on the nature of the ground state and the charge disproportionation present. For example, the experimental lower magnon bands of the layered manganite \LCom{La}{0.5}{Sr}{1.5} could be described by a localized spin model, in terms of the CE phase with valences \Mn{3} and \Mn{4}, including next-nearest-neighbor (NNN) intra-chain couplings between \Mn{4} ions~\cite{Senff}. In other cases, like for the three-dimensional \Com{Nd}{0.5}{Sr}{0.5}~\cite{Ulbrich}, the consideration of different spins, $S_1=1.79$ and $S_2=1.67$ corresponding to \Mn{3.42} and \Mn{3.66}, was found necessary to describe the measured spin-wave spectrum. More recently~\cite{Ewings}, it was concluded, after comparison with CE, ZP and two dimer phases~\cite{Sikora}, that the best fit for the measured spin excitations of the canonical three-dimensional manganite \Com{Pr}{0.5}{Ca}{0.5} corresponds to a Goodenough phase that includes NNN interactions along the chains, without charge disproportion. Notice that in Refs.\cite{Senff,Ulbrich,Ewings}) only magnon bands below $\sim$40 meV were reported, corresponding to the lower magnon branches (below the magnon excitations gap).

Concerning the bilayer compound \BCom{Pr}{Ca}{0.9}{Sr}{0.1}~\cite{TokunagaN,Tokunaga77,
Tokunaga78,Thiyagarajan,Chowdhury}, which is the central issue of our work, Johnstone~\emph{et al.}~\cite{Johnstone} presented two alternative generalized CE ground states as the most satisfactory fits which they found to their INS data. One, includes exchange couplings up to NNN within the zig-zag chains, between spins on the corner sites (as in Refs.\cite{Senff,Ulbrich}) and among the bridge sites, being corner and bridge sitesystemss occupied by equal spins~\cite{Johnstone} of magnitude $S_1=S_2=7/4$. Their alternative fit corresponds to a CE phase, without NNN couplings, but with spins of respective magnitudes $S_1=2.16$ and $S_2=1.34$, corresponding to \Mn{2.68} and \Mn{4.32} (i.e. $\Delta n=1.68e^-$): thereby assuming a charge disproportionation $\delta= 0.82$ which is outside the range $0\leq\delta\leq 0.5$ reported by most experiments. 

In this context, discussing the possibility of multiferroicity in some manganese compoundsystemss and other magnetic systems~\cite{Jardon, Mercone}, Efremov, Van den Brink and Khomskii~\cite{Efremov} proposed a new phase that could be relevant for the half-doped manganites, as it emcompasses different possibilities: the \emph{intermediate} phase (IP). This phase consists of spin dimers (thus incorporating aspects of the ZP phase~\cite{Aladine}), though formed by a pair of parallel Mn spins of different magnitude, thereby allowing for a degree of Mn charge disproportionation but not necessarily as large as the one in the original CE phase~\cite{Goodenough}. In the intermediate phase, consecutive spin dimers located along the planar zig-zag chains are oriented at a constant relative angle $\theta$ between them. Varying the charge disproportionation and the angle between the dimers, the IP phase would allow to continuously interpolate between the two limiting cases represented by: the CE phase, and the dimer phase denoted as ``orthogonal intermediate $\pi/2-$ phase"~cite{Efremov}. Some theoretical studies of the degenerate double-exchange model~\cite{Giovannetti,Picozzi} and first-principle calculation~cite{Yamauchi} gave support to the IP and found small but non-zero Mn charge disproportionations. The intermediate phase has been little explored in the context on magnon excitations in half-doped manganites~\cite{Buitrago+Ventura+Manuel}. 

In this work, we study the ground state of half-doped bilayer manganites indirectly, by calculating the magnetic excitations for the intermediate phase, including the Goodenough's CE and ZP phases. For this purpose, we extend the model analyzed in Ref.~\cite{Buitrago+Ventura+Manuel} for the layered compounds. In a localized spins scheme, we include and compare the effects, on the classical ground state and the quantum magnon spectrum, of a set of magnetic couplings, apart from those considered in Refs.~\cite{Senff,Ulbrich,Johnstone}. In particular, we study the effect of an intra-chain inter-dimer biquadratic coupling, and of a differentiation between inter-chain couplings when  the intermediate phase is characterized by non-zero $\theta $ angle between the dimers along the zig-zag chains. In our work we also introduced a new in-plane antiferromagnetic (AF) NNN coupling between inter-chain sites, which we identified as essential to describe the upper and lower magnon branches measured around the gap in bilayer manganites, concretely for \BCom{Pr}{Ca}{0.9}{Sr}{0.1}~\cite{Johnstone}. For this compound, we present two alternative fits to those proposed by Johnstone~\cite{Johnstone}, which have the advantage of not requiring NNN intra-chain couplings and, in addition, keeping the charge disproportionation within the range suggested by most experiments. In connection with our finding for bilayer compounds, we revisited the magnetic excitations of the laminar related compounds~\cite{Buitrago+Ventura}, focusing on the upper magnon branches. In this work we prove that their measurement would provide the key to identify unambiguously the ground state in layered half-doped manganites.
\section{Model for the intermediate phase, and magnon calculation method}
\label{sec.Model}
\begin{figure}[!htb]
\centering
\includegraphics[width=0.6\columnwidth]{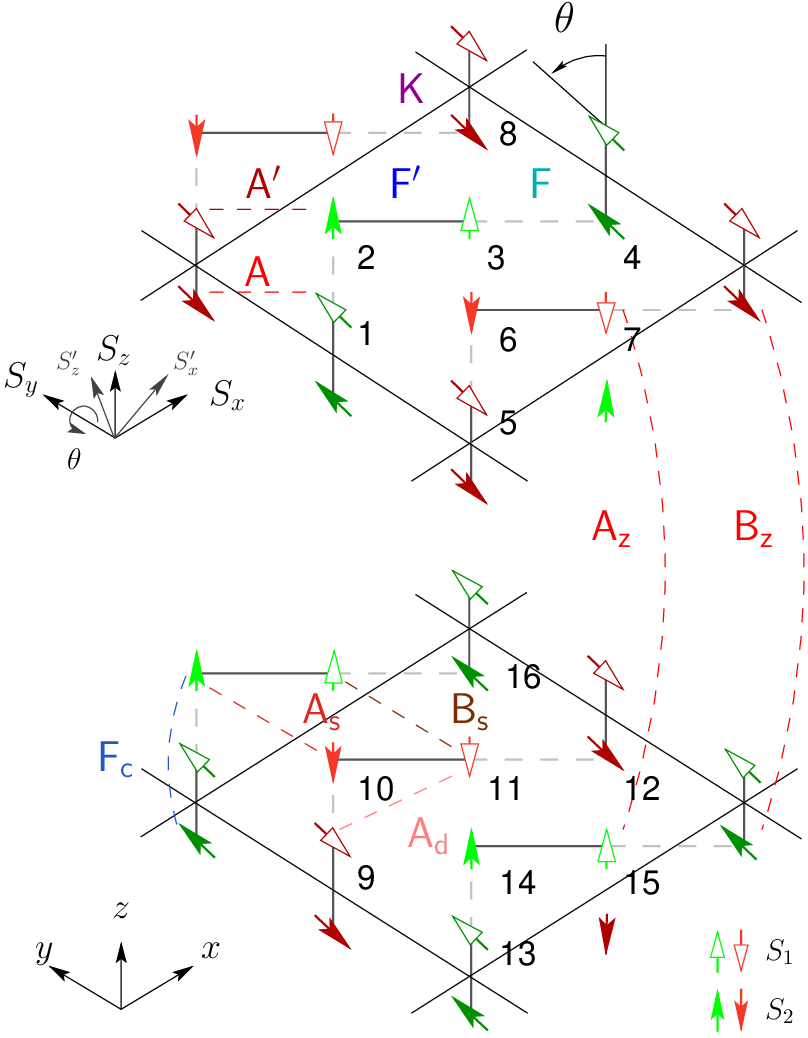}
\caption{(Color online) Schematic representation of the unit cell for the intermediate phase studied, representing by empty and filled arrows the \Mn{(3.5-\delta)} ($S_1$) and \Mn{(3.5+\delta)} ($S_2$) sites, respectively. Along the zig-zag chains, spin dimers can be identified by the solid lines representing the intra-dimer ferromagnetic coupling $\Fp$, which alternate with dashed lines representing the inter-dimer ferromagnetic coupling $\F$. The rotation of the spin quantization axes, around $S_y$, between the spins of consecutive dimers on a zig-zag chain is described by angle $\theta$. All magnetic couplings shown, are described in the text. For clarity, here the NN in-plane magnetic couplings are shown in the upper plane while the NNN are shown in the lower plane, and $\Az$, $\Bz$ represent the inter-plane couplings.}
\label{FIG-1}
\end{figure}

To describe the intermediate phase~\cite{Efremov} of half-doped bilayer manganites, we consider a model of localized spins. As for layered compounds in our previous work~\cite{Buitrago+Ventura+Manuel}, the intermediate phase
is represented by spin dimers~\cite{Aladine} formed by a pair of parallel Mn spins which could be, in principle, of different magnitude. We denote the respective spin magnitudes in a dimer by $S_1$ and $S_2$, corresponding to \Mn{(3.5-\delta)} and \Mn{(3.5+\delta)} ions: $\delta$ being a measure of the Mn charge disproportionation. In the intermediate phase, consecutive spin dimers located along the planar zig-zag chains are oriented at a constant relative angle $\theta$ between them, as shown in figure~\ref{FIG-1}. Notice that the zig-zag chains are antiferromagnetically coupled between them in the planes, as well as in the ($z$) direction perpendicular to the planes, while along $z$ equal charges (spin magnitudes) are stacked one on top of another.
 
Varying the Mn-charge disproportionation $\delta$ and the angle $\theta$, this intermediate phase would allow us to continuously interpolate between Goodenough's CE phase ($\delta=0.5$, $\theta=0$), and the dimer phase denoted as the ``orthogonal intermediate $\pi/2-$phase'' ($\delta=0$, $\theta=\pi/2$). Notice that in the latter $\pi/2-$phase~\cite{Efremov}, consecutive dimers would have their spins oriented perpendicularly, in contrast to Daoud-Aladine's ZP phase~\cite{Aladine} where all dimers in a zig-zag chain have parallel spins ($\delta=0$, $\theta=0$). We analyze these two dimer phases as well as a generalized CE phase as possible electronic ground states for the fits to the magnons measured in Pr(Ca$_{0.9}$Sr$_{0.1}$)$_2$ Mn$_2$O$_7$~\cite{Johnstone}.

Here, in order to analyze their effects and identify those most relevant to improve the description of the experimental INS results so far available in half-doped manganites, we have considered a wide range of magnetic couplings, including all those previously introduced in the fits to experiments by other groups as well as a few new additional ones, as shown in figure~\ref{FIG-1}.

Concretely, the magnetic couplings analyzed were: along the zig-zag chains two intra-chain nearest-neighbor (NN) ferromagnetic couplings, intra-dimer $\Fp$ and inter-dimer $\F$; an inter-chain NN antiferromagnetic coupling $\A$; an intra-chain next-nearest-neighbor (NNN) ferromagnetic coupling $\Fc$ 
between $S_2$ corner spins; a single-ion anisotropy $\D$ (the previous couplings were considered in the experimental fits of Refs.~\cite{Senff,Ulbrich,Johnstone}). Along $z$ we included two inter-plane antiferromagnetic couplings, allowing for eventual spin differentiation, i.e. $\Az$ between $S_1$ spins and $\Bz$ among $S_2$ spins (as for 3D manganites previously~\cite{Buitrago+Ventura}, while $\Az=\Bz$ was assumed in Refs.~\cite{Ulbrich,Johnstone}). As for layered half-doped manganites~\cite{Buitrago+Ventura+Manuel} an intra-chain NNN antiferromagnetic coupling $\Ad$ between $S_1$ bridge spins (also discussed in Ref.~\cite{Johnstone}), and an inter-dimer biquadratic (BQ) coupling term $\K$ along the zig-zag chains. As new ingredients to improve the description of experimental data, here we introduced: for the intermediate phase at non-zero $\theta$ angles, taking into account that the rotation (canting) between inter-chain spins might weaken some antiferromagnetic couplings, we introduced the possibility of a differentiation of the AF (NN) inter-chain coupling not considered before , hence we have couplings $\A$ between non-rotated dimer-spins, and $\Ap$ between spins in rotated dimers (as depicted in figure~\ref{FIG-1}).
And we also studied the effect of new NNN coupling terms between the zig-zag chains in section~\ref{sec.AsandBs}. To take into account charge disproportionation effects, we assumed inter-chain NNN couplings of magnitude $\As$ between $S_2$ spins, and $\Bs$ between $S_1$ spins. We discuss the importance of these couplings which we introduced, in section~\ref{sec.Discussion}. 

In table~\ref{TABLE1} we summarize the parameters characteristic of some specific phases proposed for half-doped manganites, in terms of the parameters of the model which would differ between them.
\begin{table}[H]
\centering
   \begin{tabular}{ccc}
   \hline\hline
    Goodenough's CE phase & Orthogonal $\pi/2$ phase & ZP dimer phase\\
    $\theta=0$ & $\theta=\pi/2$ & $\theta=0$\\
    \hline\hline
    $S_1$=2 & $S_1=1.75$ & $S_1=1.75$\\
    $S_2$=1.5 & $S_2=1.75$ & $S_2=1.75$ \\
    $\F=\Fp$ & $\F<\Fp$ & $\F<\Fp$\\
    $\Ap=\A$ & $\Ap<\A$ & $\Ap=\A$\\
	$\K=0$ & $\K\ne 0$ & $\K=0$\\
    \hline\hline
  \end{tabular}
  \caption{Model parameters characterizing specific magnetic phases.}
\label{TABLE1}
\end{table}
For a concrete implementation of the intermediate phase, able to interpolate between Goodenough's CE phase ($\delta=0.5$, $\theta=0$), and the ``orthogonal intermediate $\pi/2-$dimer phase" ($\delta=0$, $\theta=\pi/2$), we have considered the same angular dependence for $\F$, $S_1$ and $S_2$ previously used for the layered compounds~\cite{Buitrago+Ventura+Manuel}, now also including angular dependence in coupling $\Ap$, so that:
 \begin{subequations}
  \begin{gather}
   \F=\F'\,\cos\left(\frac{2\theta}{3}\right),\qquad \Ap=\A\,\cos\theta\label{EQ1.a},\\[1mm]
   S_1=\frac{7}{4}+\frac{1}{4}\,\cos\theta,\qquad S_2=\frac{7}{4}-\frac{1}{4}\,\cos\theta. \label{EQ1.b}
  \end{gather}
 \end{subequations}
Notice, that in this case: $\delta\equiv S_1-S_2=0.5\,\cos\theta$. 

With these magnetic couplings the Hamiltonian is given by: 
\begin{align}
  \begin{aligned}
    H=&-\Fp\sum_{\nn\in C, D}\Sn_i\cdot \Sn_j 
    - \F\sum_{\nn\in C\notin\,D}\Sn_i\cdot \Sn_j^{\,\prime} 
    + \A\sum_{\nn\notin C} \Sn_i\cdot \Sn_j 
    + \Ap\sum_{\nn\notin C} \Sn_i\cdot \Sn_j^{\,\prime}
    \\[2mm]&
    +\K\sum_{\nn\in C,\notin\,D} \big(\Sn_i\cdot \Sn_j^{\,\prime}\big)^2 
    - \D\sum_i \Sn_{iz}^2  
    + \Az\sum_{\nn\notin P} \Sn_i\cdot \Sn_j 
    + \Bz\sum_{\nn\notin P} \Sn_i\cdot \Sn_j 
    \\[2mm]&
    + \Ad\sum_{\nnn\in\,C} \Sn_i\cdot \Sn_j^{\,\prime}  
    - \Fc\sum_{\nnn\in\,C} \Sn_i\cdot \Sn_j^{\,\prime}
    + \As\sum_{\nnn\in\,C} \Sn_i\cdot \Sn_j 
    + \Bs\sum_{\nnn\in\,C} \Sn_i\cdot \Sn_j
  \end{aligned}
  \label{EQ2}
\end{align}
where $C$ denotes spins in one zig-zag chain, $D$ in a dimer, and $P$ in a plane, the sums over $\nn$ denote NN, while $\nnn$ refer to NNN, as usual. $\Sn_j^{\,\prime}$ is used for rotated spins.

Although, for the purposes of general discussion and comparison with previous research in the framework of the description of spin excitations in manganites, the Hamiltonian in~{\eqref{EQ2}} contains  
 12 coupling parameters, it is important to stress that in the present work we find that a  maximum of 5 non-zero couplings is needed to optimally fit the magnons.

Using the above Hamiltonian, the classical energy per unit cell of the system is written as:
\begin{align}
\begin{aligned}
	E=&8\,S_1\,S_2\Big[\K S_1 \,S_2 \cos^2\phi - \big(\A +\Fp\big) - \big(\Ap + \F\big)\cos\theta \Big]
	+ 4S_1^{\,2}\Big[2(\Ad-\Bs)\cos\theta - \Az\Big]
	\\[2mm]&
	- 4 S_2^{\,2}\Big[2(\Fc+\As)\cos\theta + \Bz \Big]
	- 4 \D\Big(S_1^{\,2}+S_2^{\,2}\Big)\Big(1+\cos^2\theta\Big)
\end{aligned}
\label{EQ3}
\end{align}
Equation~\eqref{EQ3} will allow us to evaluate the classical energy per unit cell, in order to identify the parameters which are relevant to obtain a stable classical intermediate phase, as well as discuss some classical and quantum results in next section.

To determine the quantum magnetic excitations of the intermediate phase for bilayer half-doped manganites, we considered the 16-spin magnetic unit cell shown in figure~\ref{FIG-1}. We first performed a local rotation of the spin quantization axes at an angle $\theta$, for sites within the rotated dimers in Hamiltonian~\eqref{EQ2}. The spins in the rotated system $\Sn_j^{\,\prime}$ are related to the fixed system ones $\Sn_j$ as follows: $S_j^{\,y\prime}=S_j^{\,y}$, 
\begin{subequations}
 \begin{gather}
  S_j^{\,x}=S_j^{\,x\prime}\,\cos\theta - S_j^{\,z\prime}\,\sin\theta\, ,\label{EQ4.a} \\
  S_j^{\,z}=S_j^{\,x\prime}\,\sin\theta + S_j^{\,z\prime}\,\cos\theta\, .\label{EQ4.b} 
 \end{gather}
\end{subequations}
Afterwards, we use the Holstein-Primakoff transformation for spin operators, Fourier transform, and determine the magnon excitations by paraunitary diagonalization~\cite{Colpa} of the Hamiltonian matrix in the linear spin waves approximation, as in~\cite{Buitrago+Ventura+Manuel}.

\section{Results: effects of different coupling parameters.}
\label{sec.Results}
Below, we present the most relevant results obtained in our classical and quantum investigation of the intermediate phase. We analyse, in particular, the effects of the different magnetic coupling parameters included in our model, allowing us to identify which parameters are most relevant to stabilize the magnetic excitations of the intermediate phase as a means of indirectly analizing the stability of that proposed ground state, and which parameters enable us to exhibit an improved fit of the inelastic neutron scattering data for bilayer compounds reported in the literature~\cite{Johnstone}. In the following, all energy parameters mentioned will be expressed in meV, unless explicitly stated otherwise.
\subsection{Effect of intra-chain inter-dimer biquadratic coupling $\K$}
As mentioned in previous section, in our model we included biquadratic coupling $\K$, which was originally used by Cieplak~\cite{Cieplak} to study magnetic spins in mixed-valence three-dimensional manganites. Also, in a series of recent studies~\cite{wysocki,stanek,chaloupka} of the spin dynamics exhibited by ferropnictide superconductors, biquadratic exchange terms were invoked as necessary ingredients to be added to an anisotropic Heisenberg Hamiltonian in order to describe the experimental results.\cite{diallo,zhao} We have studied the effect of $\K$ previously, in the context of our investigation of the magnons of layered half-doped manganites and the intermediate phase~\cite{Buitrago+Ventura+Manuel}. 

\begin{figure}[H]
 \centering
 \includegraphics[width=0.6\columnwidth]{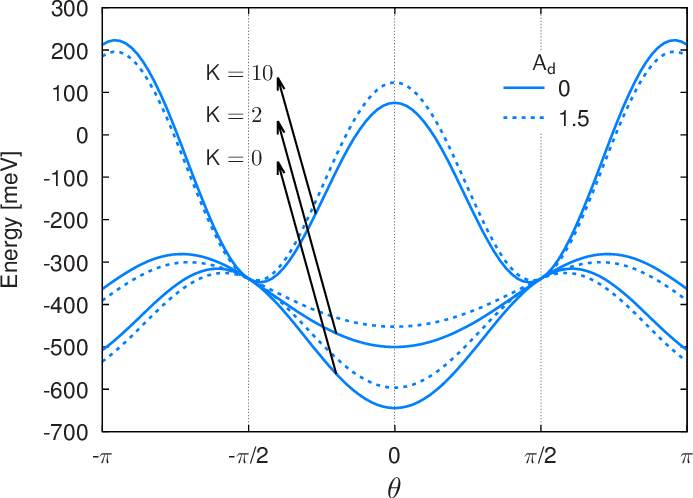}
 \caption{(color online) Bilayer: Angular dependence of the energy of the classical intermediate phase. Non-zero magnetic couplings (in meV): $\Fp=11.39$, $\A=1.5$, $\D=0.074$, $\Az$($=\Bz$)$=0.88$ (as in Ref.\cite{Johnstone}). $\F$ as in equations~\eqref{EQ1.a} and $S_1$, $S_2$ as in equations~\eqref{EQ1.b}. $\Ad=0$ (solid lines), $\Ad=1.5$ (dashed lines). The $\K$ dependence is indicated by arrows (for each solid line, and its adjacent dashed line).}
 \label{FIG-2}
\end{figure}

As shown in figure~\ref{FIG-2}, this magnetic coupling allows to obtain classically stable intermediate phases with $\theta$ angles between 0 and $\pi/2$. $\K$ is the most relevant coupling for the stabilization of the classical intermediate phase without requiring the introduction of other NNN couplings, since it favors perpendicular arrangement of inter-dimer adjacent spins. Although in our model, to simplify the quantum magnon calculation, we introduced $\K$ only between dimers within a zig-zag chain, we checked that one could extend it to inter-chain neighbor sites without substantially altering its effect on the stabilization of the classical intermediate phase. 

With respect to the layered case~\cite{Buitrago+Ventura+Manuel}, the main difference lies in the energy scales involved. Having only half of the $\K$ couplings, and $\Az$($=\Bz$)$=0$, the range of values for the classical energy of the intermediate phase of layered compounds extends between -320 y 120 meV, in contrast to the range of 650 a 270 meV shown for bilayer compounds in figure~\ref{FIG-2}. We have verified that the angle $\theta_{\text{min}}$ (by which we denote the angle characteristic of the classical intermediate phase with minimal energy, which in general depends on $\K$ as e.g. figure~\ref{FIG-2} shows), for each particular value of $\K$ is the same regardless of whether the system is layered or bilayered.
\subsection{Differentiation of inter-chain couplings $\A$ y $\Ap(\theta)$}
\label{sec.AandAp}
\begin{figure}[H]
\centering
\includegraphics[width=0.6\columnwidth]{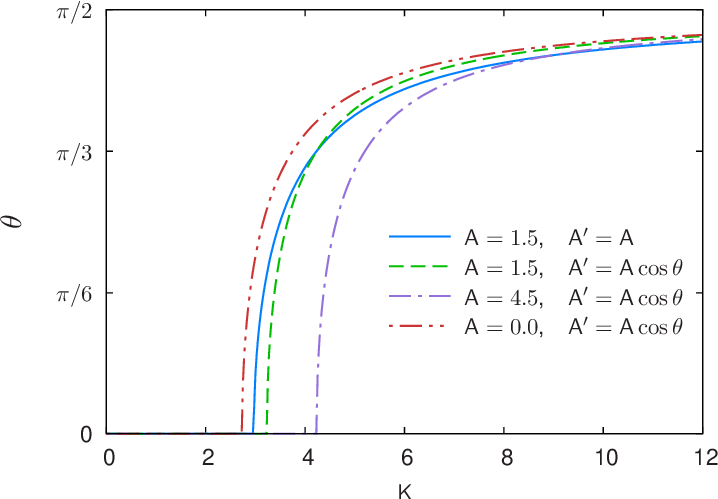}
\caption{(color online) Bilayer: Angle characteristic of the classical intermediate phase with minimal energy, as a function of $\K$: for various $\A$ and $\Ap$ inter-chain couplings, as detailed in the inset. Non-zero magnetic couplings, in meV: $\Fp=11.39$, $\Az$($=\Bz$)$=0.88$, $\D=0.074$. $\F$ as in equations~\eqref{EQ1.a} and $S_1$, $S_2$ as in equations~\eqref{EQ1.b}. For: $\A=\Ap=1.5$ (solid line); $\A=1.5$, $\Ap=1.5\,\cos\theta$ (dashed line); 
 $\A=4.5$, $\Ap=4.5\,\cos\theta $ (dash-dotted line); $\A = 0 = \Ap$ (dash-double dotted line).}
\label{FIG-3}
\end{figure}

As seen in figure~\ref{FIG-3}, including a differentiation in the antiferromagnetic coupling between chains, as described in section~\ref{sec.Model}, increases  the critical value of biquadratic coupling $\Kc$ above which a classical intermediate phase with $\theta \neq 0$ is stable. In particular, we find that even for $\A=0$ a critical value $\Kc=2.75$ is necessary to obtain a stable intermediate phase with non-zero angle, feature which we find is common to the layered and bilayer cases. Indeed using for $\Ap$ the angular dependence $\Ap=\A\,\cos\theta$, an increase of angle $\theta$ results in a change of the effective AF inter-chain coupling which reduces the energy of the system especially for $\theta\sim0$, which in turn makes it more difficult to stabilize an intermediate phase with larger $\theta$.

\begin{figure}[H]
 \centering
 \includegraphics[width=0.6\columnwidth]{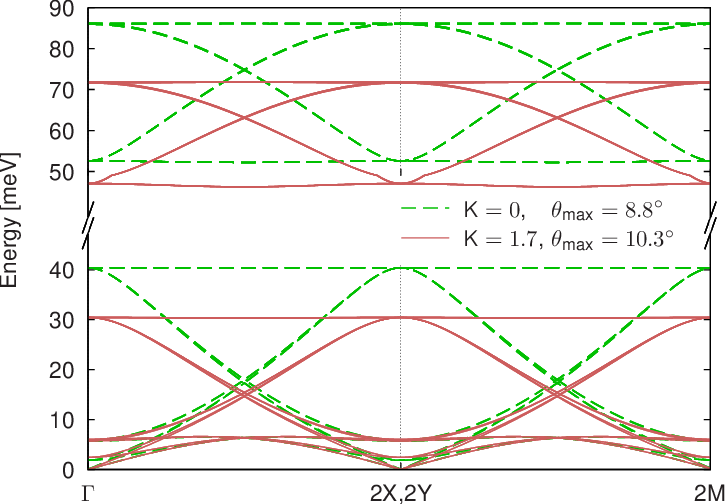}
 \caption{(color online) Bilayer: intermediate phase quantum magnons, for different $\K$. Non-zero magnetic couplings, in meV: $\Fp=11.39$, $\A=1.5$, $\D=0.074$, $\Az$($=\Bz$)$=0.88$; $\F$ and $\Ap$ as in equations~\eqref{EQ1.a}, $S_1$, $S_2$ as in equations~\eqref{EQ1.b}. For: $\K=0$ and $\theta=\theta_{\text{max}}=8.83$ degrees (dashed lines); $\K=1.7$ and $\theta=\theta_{\text{max}}=10.39$ degrees (solid lines).}
 \label{FIG-4}
\end{figure}

Considering now the magnons of the quantum intermediate phase presented in figure~\ref{FIG-4}, for couplings $\A=1.5$ meV and \mbox{$\Ap=\A\cos\theta$} corresponding to the classical case represented by the dashed-line of figure~\ref{FIG-3}, it is evident that the effect of $\K$ is to widen the range of $\theta$ angles which lead to a stable quantum intermediate phase. In particular, the maximum angle $\theta_{max}$ below which a stable quantum intermediate phase exists is seen to shift from $8.83$ degrees at $\K=0$, to $10.39$ degrees at $\K=1.7$. Unless otherwise stated, all quantum magnon results appear plotted along symmetry paths in the square or cubic lattice Brillouin zones, depending on the case. Notice also that for the bilayer compounds we are using the same notation for the BZ of Ref.\cite{Johnstone}, since to take into account twinning effects it is necessary to double the unit cell in real space, so the points in reciprocal space here labelled as I and II, are given respectively by: I by X (or Y), and II by $(2\pi/a_x,\pi/a_y,0)$ (or $(\pi/a_x,2\pi/a_y,0)$), where $a_i$ denotes the lattice parameter along direction $i (i = x,y)$. Comparing figure~\ref{FIG-3} and figure~\ref{FIG-4} it is clear that in both cases, $\K=0$ or $\K=1.7$, the quantum fluctuations lead to stable quantum intermediate phases with $\theta\neq 0$ in contrast with the corresponding stable classical phase, having $\theta=0$. But it is worth mentioning that, using the same in-plane coupling parameters, we find that essentially the same $\theta_{max}$ value is obtained for the layered and the bilayer cases, in agreement with the classical result mentioned in previous subsection.

\subsection{Effect of long-range inter-chain couplings: $\As$ and $\Bs$}
\label{sec.AsandBs}
We refer the reader to \ref{sec.FcandAd}, for a thorough discussion of the effects of the magnitude and sign of the intra-chain NNN couplings along the zig-zag chains $\Fc$ and $\Ad$, in our notation of figure~\ref{FIG-1} and equation~\eqref{EQ2}. We show that, in contrast to suggestions made in previous works~\citep{Senff}, the effects of these two couplings are noticeably different not only on the classical energy (and the resulting stable intermediate phase) but even more so on the quantum magnons, where different magnon branches are affected by each  NNN intra-chain exchange interaction. In particular, in the \ref{sec.FcandAd} we show that $\Fc$ and $\Ad$ are respectively relevant to determine the top and bottom of the gap in the magnon spectrum of the bilayer manganite. 

Here, we discuss the effect of  NNN coupling terms between the zig-zag chains which, to our knowledge, have not been included before in fits to INS results for half-doped manganites. In the following we will show our findings, in particular how each of them affects the different magnon branches. In section~\ref{sec.Discussion} we will discuss their relevance to obtain an improved fit for the magnetic excitations measured in \BCom{Pr}{Ca}{0.9}{Sr}{0.1}~\cite{Johnstone}. Concretely, we decided to include these inter-chain NNN coupling parameters in the model, to be able to take into account charge disproportionation effects: $\As$ between $S_2$ spins, and $\Bs$ between $S_1$ spins. 

\begin{figure}[H]
 \centering
 \includegraphics[width=0.6\columnwidth]{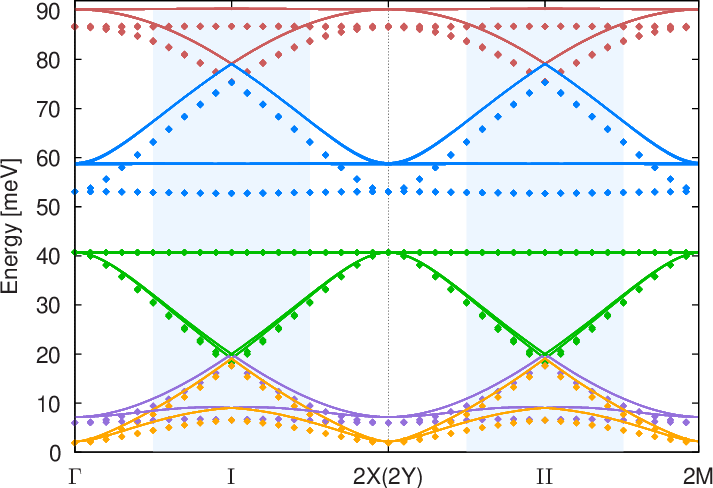} 
 \caption{(color online). Bilayer: Effect of $\As$ on the quantum intermediate phase magnons. Non-zero magnetic couplings, in meV: $\Fp$(=$\F$)$=11.39$, $\A(=$\Ap$)=1.5$, $\Az$($=\Bz$)$=0.88$, $\D=0.074$; while: $\As=0$ (symbols), and $\As=2.0$ (solid lines). $S_1=2$, $S_2=1.5$. $\theta = 0.$}
 \label{FIG-5}
\end{figure}
As shown in figure~\ref{FIG-5}, the main effect of non-zero antiferromagnetic $\As $ couplings between $S_2$ spins is to increase almost rigidly the energies of all upper magnon branches (those above the magnon gap), while the only most relevant change to energies of lower magnon branches (those below the magnon gap) is produced around BZ points I and II (lifting the lowest-energy magnon branch about 2 meV, at I and II).

\begin{figure}[H]
 \centering
 \includegraphics[width=0.6\columnwidth]{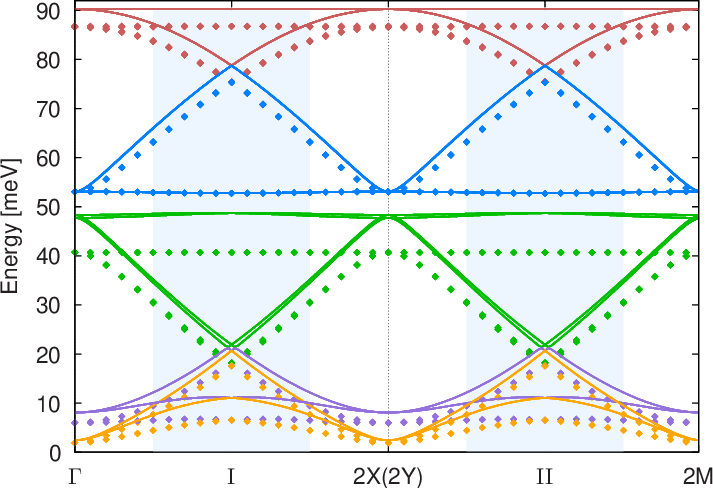}
 \caption{(color online). Bilayer: Effect of $\Bs$ on the quantum intermediate phase magnons. Non-zero magnetic couplings, in meV: $\Fp$(=$\F$)$=11.39$, $\A(=$\Ap$)=1.5$, $\Az$($=\Bz$)$=0.88$, $\D=0.074$; while: $\Bs= 0$ (symbols), and $\Bs= 2.0$ (solid lines). $S_1=2$, $S_2=1.5$. $\theta = 0.$}
 \label{FIG-6}
\end{figure}
As shown in figure~\ref{FIG-6} (for antiferromagnetic $\Bs>0$), the effect of $\Bs$ coupling between antiparallel $S_1$-$S_1$ spins is more complex (than that of $\As$), starting by the q-dependent shifts in the magnon branches. For instance, the lowest magnon branch, has a maximum increase in energy at BZ points I and II, while the magnons at $\Gamma, 2X, 2M$ are unaffected. The largest magnon change resulting from $\Bs > 0$ is observed at the flat band corresponding to the bottom of the magnon gap, which is lifted rigidly by about 8 meV in figure~\ref{FIG-6}. Changing the sign of this coupling: to ferromagnetic $\Bs<0$, results in a reversion of the effects on the magnons: the ensuing q-dependent shift of energies has opposite sign (and equal magnitude). Notice that the magnon branch at 60 meV is unaffected along the whole BZ by the appearance of a non-zero $\Bs$ coupling.

\section{Discussion: comparison with experiments, and predictions.}
\label{sec.Discussion}
In this section, we will show some results which allow us to discuss and contribute to clarify a few important points relative to the nature of the ground state of half-doped manganites, still unresolved, which we can address using our model. 

\subsection{\textbf{Improved fits for {\rm\bf \BCom{Pr}{Ca}{0.9}{Sr}{0.1}} magnons.}}

\begin{figure}[H]
 \centering
 \includegraphics[width=0.6\columnwidth]{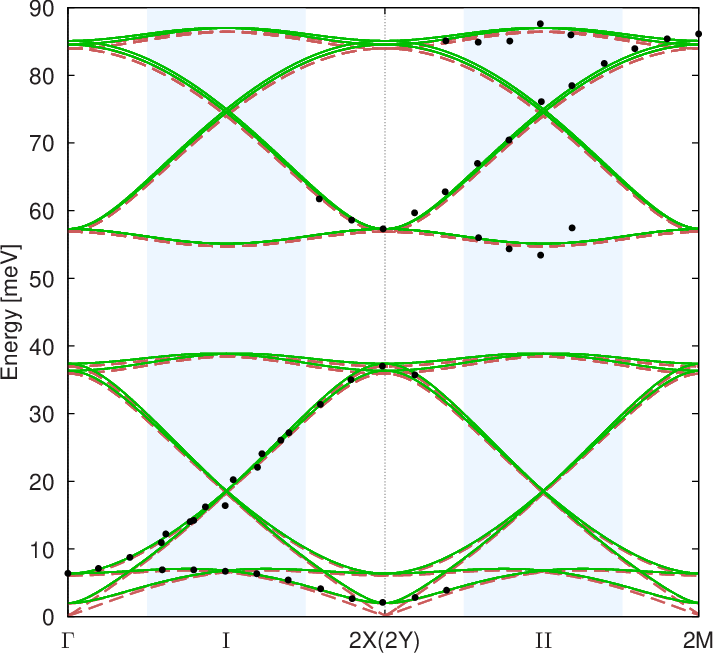}
 \caption{(color online) Bilayer: fits to the INS data for magnons~\cite{Johnstone} (included as black dots), using an intermediate phase including NNN couplings between zig-zag chains, but not along the chains. Non-zero magnetic couplings, in meV: $\Fp=10.89$, $\A=3.6$, $\Az$($=\Bz$)$=0.88$, $\As=0.5$, $\Bs=-3.4$, $\D=0.074$, $\F$ and $\Ap$ as in equation~\eqref{EQ1.a}. For these phases, we did not use equations~\eqref{EQ1.b} for the spin magnitudes, instead: $S_1=1.8$ and $S_2=1.7$ (i.e. $\delta=0.2 $). $\theta=8$ degrees (dashed lines). Optimal fit: $\theta=0$ (solid lines).}
 \label{FIG-8-n}
\end{figure}

For bilayer \BCom{Pr}{Ca}{0.9}{Sr}{0.1}~\cite{Johnstone}, we will now present two other possible fits for the INS measurements by Johnstone\cite{Johnstone}. In particular, the second one presented  in figure~\ref{FIG-8-n} is the optimal fit to the experimental results yet obtained, to our knowledge.

\begin{figure}[H]
 \centering
 \includegraphics[width=0.6\columnwidth]{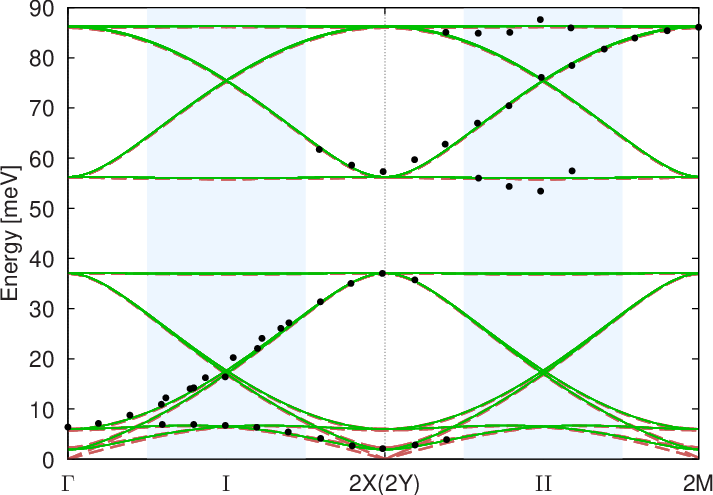}
 \caption{(color online) Bilayer: fits to the experimental magnon data~\cite{Johnstone} (included as black dots) in terms of an intermediate phase with non-zero biquadratic coupling $\K$, describing the data as the fit by Johnstone~\emph{et al.}~\cite{Johnstone}. Non-zero couplings, in meV: $\Fp=13.45$, $\A=1.5$, $\Az$($=\Bz$)$=0.88$, $\K=1.7$, $\D=0.08$, $\F$ and $\Ap$ as in Eq.~\ref{EQ1.a}. Comparison of two fits $\theta=0$, $S_1=2$ and $S_2=1.5$ (solid line); $\theta=9.4$ degrees, $S_1$ and $S_2$ as in equation~\eqref{EQ1.b} (dashed line).}
 \label{FIG-7-n}
\end{figure}

First, in figure~\ref{FIG-7-n} we present a fit to the magnons measured in \BCom{Pr}{Ca}{0.9}{Sr}{0.1}~\cite{Johnstone} which we obtained using our model with $\K\ne 0$ and $\theta=0$ (solid line). Except for the inclusion of $\K$, the coupling magnitudes here used are quite similar to those used by Johnstone~\emph{et al.}~\cite{Johnstone}, i.e. the ratios between the couplings are similar. As is seen, our fit to the experimental data (shown in solid-line in figure~\ref{FIG-7-n}) is almost indistinguishable from their own fit. Nevertheless, the present fit has the advantage that it involves a charge disproportionation ($\delta=0.5$: thus \Mn{3} and \Mn{4} would be present) within the range proposed by Goodenough~\cite{Goodenough} and most other experiments~\cite{Hill,Mercone}, even without including NNN couplings like $\Ad$ or $\Fc$, as proposed in Refs.~\cite{Senff,Ulbrich,Johnstone} for their best fits. We use $\theta$=0, which is the angle corresponding to the minimum of the classical energy for the intermediate phase, with these parameters, because, as we shown in figure~\ref{FIG-7-n}, when we use the maximun angle to obtain a stable intermediate phase, $\theta=9.4$, the lowest magnon branches differ from experimental magnon data.

Finally, we present the best fit which we obtained for the INS data for magnons in the half-doped bilayer manganite \BCom{Pr}{Ca}{0.9}{Sr}{0.1}~\cite{Johnstone} after our thorough exploration of parameter ranges and different phases for the localized spin model here studied. In figure~\ref{FIG-8-n} we exhibit the calculated magnons, comparing them with the experimental data by Johnstone~\emph{et al.}~\cite{Johnstone}. An improvement with respect to previous fits, is given by the fact that we are also able to describe the dispersion in the shaded region near II which is clearly present in the measured upper magnon bands (above the gap). We find this is due to our inclusion of NNN inter-chain couplings $\As$ and $\Bs$ not considered in previous fits, and their delicate interplay with the value of NN inter-chain coupling $\A$ for the determination of the gaps and dispersion of the bands near I and II. Our best fit to the experimental data in half-doped bilayer compounds~\cite{Johnstone} appears represented by solid lines in figure~\ref{FIG-8-n}, and corresponds to a generalized CE phase with $\theta=0$, and the following non-zero magnetic couplings (in meV): $\Fp $($=\F$)$=10.89$, $\A$($=\Ap$)$=3.6$, $\Az$($=\Bz$)$=0.88$, $\As=0.5$, $\Bs=-3.4$ and $\D=0.074$. For this particular phase, we did not use equations~\eqref{EQ1.b} for the spin magnitudes, instead we found that we optimized our fit with: $S_1=1.8$ and $S_2=1.7$ (i.e. $\delta=0.2$). In this case the frustration, due to the $\Bs$ coupling between \Mn{3} sites in consecutive chains, is similar to that proposed by Johnstone~\cite{Johnstone} with $\Ad$ but has different effects on the energies of the magnons as seen in figure~\ref{FIG-6}. As in figure~\ref{FIG-7-n}, we have seen that considering $\theta=8$ degrees, again affects the lowest magnon branches, separating these from the experimental magnon data, more noticeably near  the $\Gamma$ and $2X$ points.

\subsection{\textbf{Layered manganites}}
Motivated by the optimal fit for magnons in bilayered manganites discussed in section 4.1, we revisited the magnetic excitations of the single-layer related manganites, focusing now on the upper magnon branches (not discussed in our previous work~\citep{Buitrago+Ventura},  nor measured yet). 

\begin{figure}[H]
 \centering
 \includegraphics[width=0.6\columnwidth]{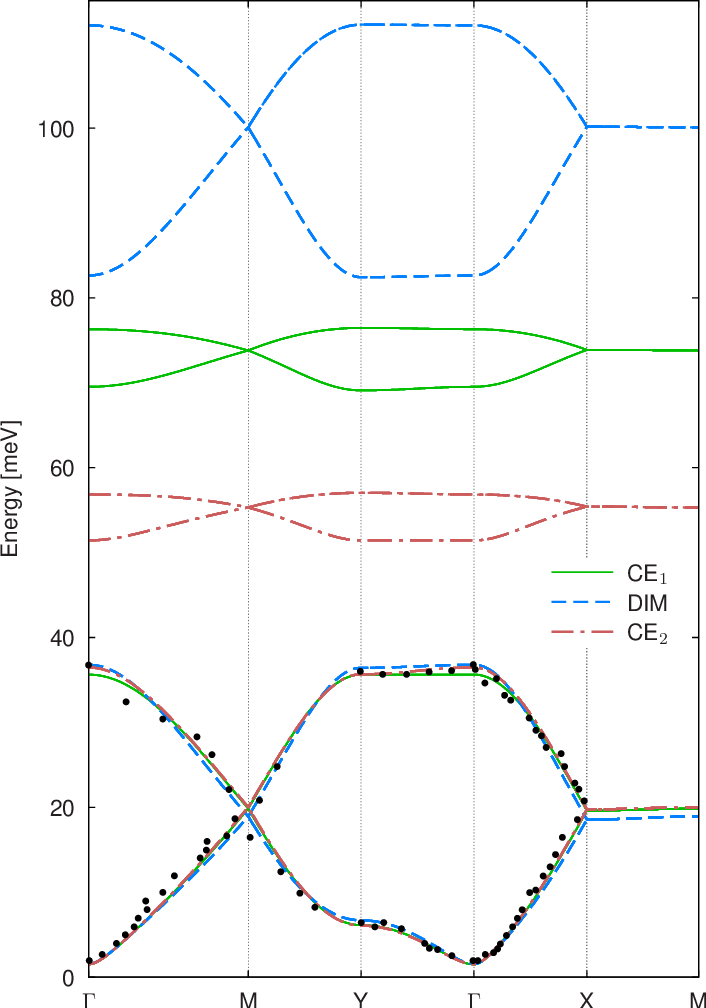}
 \caption{(color-online) Quantum magnons for layered \LCom{La}{0.5}{Sr}{1.5}. Non-zero magnetic couplings (in meV) and spin magnitudes for three proposed phases, respectively: CE$_1$ phase (solid lines): $\Fp$($=\F$)$=9.98$, $\A$($=\Ap$)$=1.83$, $\Fc=3.69$, $S_1=2$, $S_2=1.5$, fit by Senff \emph{et at.}~\cite{Senff}; ZP dimer-phase (dashed lines): $\Fp=21.58$, $\F=8.48$, $\A$($=\Ap$)$=1.98$, $S_1=S_2=1.75$; CE$_2$ phase (dash-dotted lines): $\Fp$($=\F$)$=7.18$, $\A$($=\Ap$)$=1.9$, $\Ad=-3.0$, $S_1=2$, $S_2=1.5$; INS experimental points ~\cite{Senff} (black dots). In all cases $\theta=0$.}
 \label{FIG-9-n}
\end{figure}

Concretely, for layered \LCom{La}{0.5}{Sr}{1.5}~\cite{Senff}, we compare the magnons corresponding to different assumptions for the ground state, all characterized by $\theta=0$, and zero inter-plane coupling parameters ($\Az=\Bz=0$). It is important to stress that only measurements for the lower magnon bands, below the magnon gap, were reported. Based on this, it was indicated that a fully satisfactory fit was attained assuming that the ground state was Goodenough's CE phase of table~\ref{TABLE1} (with parameters: $\Fp=\F$=9.98, $\A=\Ap=1.83$) complemented by the addition of a ferromagnetic coupling $\Fc=3.69$ between NNN intra-chain \Mn{4}-\Mn{4} pairs~\cite{Senff}. We will denote this phase as CE$_1$, in the following, and in figure~\ref{FIG-9-n} have plotted as solid lines the magnons (lower and also upper bands) corresponding to this phase, while the black dots reproduce the experimental data~\cite{Senff}. Next, we address the possibility, mentioned by Senff~\emph{et al.}, of obtaining an equally good fit for those experimental data in terms of a ground state which we will denote CE$_2$. This one consists of Goodenough's CE phase of table~\ref{TABLE1} complemented by the addition of a ferromagnetic coupling $\Ad$, between NNN intra-chain \Mn{3}-\Mn{3} pairs. In~\ref{sec.FcandAd} we analized and compared the separate effects of $\Fc$ and $\Ad$ and showed that these two coupling parameters produced different quantum phase magnons, even if one chose them so as to preserve the scaling relation leading to equal classical results using one or the other (see figures~\ref{FIG-10-n} and~\ref{FIG-11-n}). Thus, to achieve our best fit to the experimental INS data of Senff~\emph{et al.} in terms of a CE$_2$ phase, included in figure~\ref{FIG-9-n} using dot-dashed lines, diffferent magnitudes for the coupling parameters were needed: $\Fp$($=\F$)$=7.18$, $\A$($=\Ap$)$=1.9$, $\Ad=-3.0$. Finally, in figure~\ref{FIG-9-n} we also included our best fit to the experimental INS data~\cite{Senff} in terms of ZP-dimer phase~\cite{Aladine}, like indicated in table~\ref{TABLE1}, in which case: $\Fp=21.58$, $\F=8.48$, $\A$($=\Ap$)$=1.98$.

Figure~\ref{FIG-9-n} illustrates two points very clearly: {\bf (i)} knowledge limited to the lower magnon bands (below the gap) of the layered half-doped manganites is insufficient to determine unambiguously if their ground state corresponds to a CE (be it CE$_1$, or CE$_2$) or a ZP dimer-phase. We show that it is possible to attain equally good fits to the magnon branches below the gap with these three proposed ground states. The fits mainly differ (and only slightly) along the \textsf{X-M} path in the BZ, where there are no experimental data~\cite{Senff}; {\bf (ii)} it is the pending measurement of the upper magnon bands (above the magnon gap) which would unambiguously allow to recognize if the nature of the ground state of the layered half-doped manganites is CE or ZP-dimer like. This conclusion is in agreement with what Johnstone~\emph{et al.}~\cite{Johnstone} found for half-doped bilayer compounds.
 
\section{Conclusions}
We studied the spin excitations of a model of interacting localized spins suitable to describe the intermediate phase of half-doped bilayer manganites such as \BCom{Pr}{Ca}{0.9}{Sr}{0.1}, for which experimental inelastic neutron scattering data were obtained~\cite{Johnstone}. The model investigated includes magnetic couplings between Mn-ions up to second neighbors in the zig-zag chains, as well as between zig-zag chains, studied here for the first time. This allowed us to compare in detail the main effects of the different couplings on the magnons, including all the previously used to fit experimental INS data in half-doped manganites by other groups. The model we proposed for the intermediate phase~\cite{Efremov}, allows us to discuss different degrees of Mn-charge disproportionation, as well as different relative orientations of the consecutive spin dimers along a zig-zag chain. The model allows to describe Goodenough's original~\cite{Goodenough} and also generalized CE phases, the orthogonal intermediate-$\pi/2$ dimer phase~\cite{Efremov}, and the Zener-polaron dimer phase~\cite{Aladine}.

For classical spins, we find that an intermediate phase characterized by non-zero angle $\theta$ is stable only for biquadratic inter-dimer couplings above a critical value $\Kc$, which increases when increasing the coupling between the zig-zag chains ($\A$ and $\Ap$). For quantum spins, the stability of the intermediate phase is reduced to a more limited range of angles, being $\K$ the most relevant parameter for stability. In particular, we could not find any set of parameters for which the quantum $\theta=\pi/2$ orthogonal dimer-phase~\cite{Efremov} was stable. Calculating the quantum magnons of the intermediate phase, we could also analize those corresponding to Goodenough's original and generalized CE phases, and we also studied the Zener-polaron dimer phase, discussing our results in the context of recent experimental results for layered and bilayer half-doped manganites. 

We obtain an excellent agreement between theory and the magnons reported for half-doped bilayer manganites~\cite{Johnstone} when the ground state is assumed to be a generalized Goodenough's CE phase $\theta=0$, with a Mn-charge disproportionation inside the experimentally expected range ($\delta=0.2 \leq 0.5$)  in agreement with most of the experimental evidence and in contrast to Johnstone~\emph{et al.}~\cite{Johnstone} which in their fit had $\delta = 0.82$ (i.e.: \Mn{2.68} and \Mn{4.32}).  We also found that the next-nearest-neighbour magnetic couplings introduced between the planar Mn zig-zag chains, a new ingredient with respect to previous fits, and an appropriate tuning of the nearest-neighbour inter-chain antiferromagnetic coupling, are the relevant factors needed to reproduce the dispersion exhibited by the measured upper and lower magnon branches around the gap, which has not been described previously~\cite{Johnstone}. 

In connection with this finding, we revisited the magnetic excitations of the laminar related compounds~\cite{Buitrago+Ventura+Manuel}, now focusing on the upper magnon branches. Here  we predict  that important differences would appear in the yet unmeasured magnon bands above the gap between CE phases and the Zener-polaron dimer phase, which would allow to unambiguously distinguish between them. Meanwhile the measured lower magnon bands can be almost equally well described by any of the proposed phases. Therefore, at least on the basis of the available magnon data for layered half-doped manganites, one can not exclude any of the proposals for the ground state, in particular a ZP-dimer phase. Based on our predictions of Figure~\ref{FIG-9-n}, this would require INS measurements in layered half-doped manganites in the range of energies above 40 meV, similar to the range where the bilayer upper magnon bands were measured~\cite{Johnstone}.

One of the motivations of ours work, here for bilayer compounds and in our first study of Ref.~\cite{Buitrago+Ventura+Manuel} for layered ones, has been to explore if the intermediate phase proposed by Efremov, Van den Brink and Khomskii~\cite{Efremov} for manganites, might provide a better description for the experimental inelastic neutron scattering data in these compounds. In particular, if allowing for the possibility of non-parallel orientation (non-zero angle $\theta$)  between consecutive dimers along a zig-zag planar chain would improve the description of magnetic excitations. As a result of the calculations of the spin excitations for the intermediate phase one conclude that the optimal description of the available experimental results is obtained by assuming that no rotation is present between consecutive dimers along the zig-zag chains. 

Regarding the three-dimensional manganite \Com{Pr}{0.5}{Ca}{0.5}, it would be interesting to calculate the magnetic structure factor $S(\textbf{q}, \omega)$ considering the NNN magnetic couplings between chains, which we found improve the fit of the magnetic excitations of the bilaminar manganite \BCom{Pr}{Ca}{0.9}{Sr}{0.1} and were proposed for the first time in the present work.  And in particular check  if an improved fit may be obtained  including a non-zero charge disproportionation between Mn ions, as several experiments suggested and in contrast with the best fit of Ref.~\cite{Ewings}.

\section{Acknowledgements}
We acknowledge financial support by CONICET (PIP grant 0702, and the fellowship awarded to I.R.B.). L.O.M. and C.I.V. are members of Carrera del Investigador Cient\'{\i}fico, CONICET.

\appendix
\section{Effects of intra-chain NNN couplings $\Fc$ and $\Ad$}
\label{sec.FcandAd}

\subsection{Layered compounds.}
\begin{figure}[H]
 \centering
 \includegraphics[width=0.6\columnwidth]{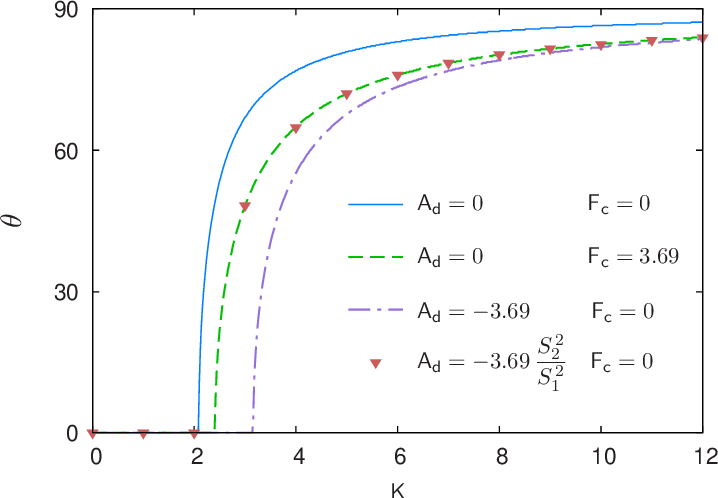}
 \caption{(color online) Layered: Angle characteristic of the classically stable intermediate phase, for different $\Ad$ and $\Fc$ combinations. Other non-zero magnetic couplings, in meV: $\Fp=9.98$, $\A=1.83$, $\D=0.05$ (as in Ref.\cite{Senff}); $\F$ and $\Ap$ as in equations~\eqref{EQ1.a}, $S_1$, $S_2$ as in equations~\eqref{EQ1.b}. For: $\Fc=0$, $\Ad=0$ (solid line); $\Fc=3.69$, $\Ad=0$, as in Ref~\cite{Senff} (dashed line); $\Fc=0$, $\Ad=-3.69$ (dash-dotted line); $\Fc=0$, $\Ad=-3.69\,S_2^{\,2}/S_1^{\,2}$ (triangles).} 
 \label{FIG-10-n}
\end{figure}

For \LCom{La}{0.5}{Sr}{1.5}, Senff~\emph{et al.}~\cite{Senff} discussed the inclusion of NNN ferromagnetic couplings along the zig-zag chains: i.e. $\Fc$ and $\Ad$ in our notation of figure~\ref{FIG-1} and equation~\eqref{EQ2}. They mention that both ferromagnetic couplings improve the fit of the magnetic excitations measured, favouring the addition of a non-zero $\Fc$ coupling to the basic parameters describing the CE phase. Our figure~\ref{FIG-10-n} shows that the effect of adding one or the other of these two couplings (with signs corresponding to ferromagnetic exchanges, for both) on the classical energy (and the resulting stable intermediate phase) is quantitavely different. For example, different values of $\Kc$ result if non-zero $\Ad$ or $\Fc$ of the same magnitude are included. This is due to the fact that the contribution to the classical energy by terms proportional to $\Ad$ is weighed by $S_1^{\,2}$, whereas the contributions proportional to $\Fc$ appear weighed with $S_2^{\,2}$, where $S_1\ge S_2$ if charge disproportionation is present. In fact, in figure~\ref{FIG-10-n}. we show how the magnon curves (line labelled by triangles, and dashed line) collapse together if these coupling parameters were accordingly rescaled: $\Ad= - \Fc S_2^{\,2}/S_1^{\,2}$.

\begin{figure}[H]
 \centering\includegraphics[width=0.6\columnwidth]{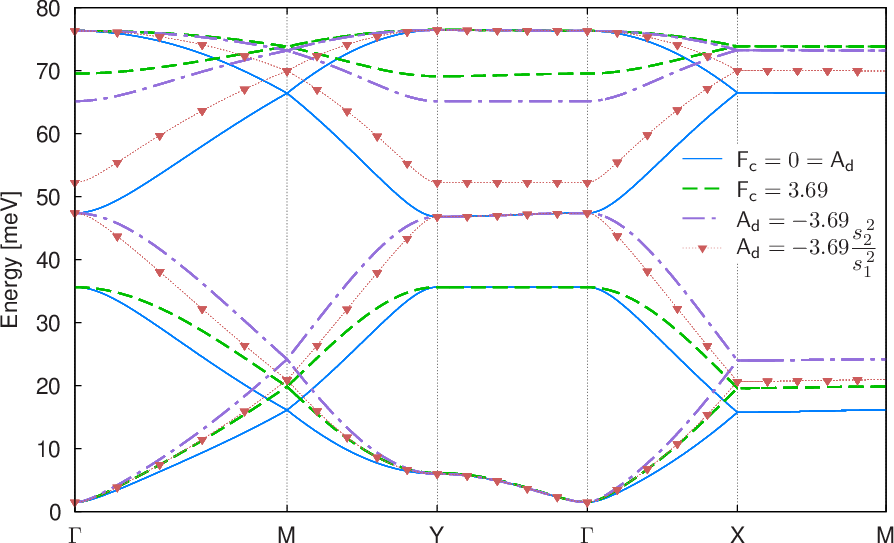}
 \caption{(color-online) Layered: Quantum intermediate phase magnons, for different $\Ad$ and $\Fc$ combinations (both considered FM). Other non-zero magnetic couplings, in meV: $\Fp (=\F) = 9.98$, $\A (= \Ap )= 1.83$, $\D= 0.05$, while spin magnitudes: $S_1=2$, and $S_2=1.5$ (as in Ref~\cite{Senff}). For: $\Fc=0$ and $\Ad=0$ (solid lines); $\Fc=3.69$ and $\Ad=0$ (dashed lines); $\Fc=0$ and $\Ad=-3.69$ (dash-dotted lines); $\Fc=0$ and $\Ad=-3.69\,S_2^{\,2}/S_1^{\,2}$ (triangles). $\theta = 0.$}
 \label{FIG-11-n}
\end{figure}

In the quantum case, the effects of $\Ad$ and $\Fc$ differ even more: along the Brillouin zone, differences in the magnons are present even for cases where the classical curves collapse together, as we show in figure~\ref{FIG-11-n}. While $\Ad$ increases the energies of the excitations in all paths of the Brillouin zone, marked differences are found for corresponding non-zero $\Fc$: in particular, for the excitations above 20 meV along the $Y-\Gamma$,$M-Y$ and $\Gamma-X$ paths. Furthermore, the magnon gap obtained with non-zero $\Fc$ is about 7 times larger than the one resulting if the correspondingly scaled $\Ad$ value (i.e. leading to equal classical energies) is used.

\subsection{Bilayer compounds.}
\begin{figure}[H]
 \centering
 \includegraphics[width=0.6\columnwidth]{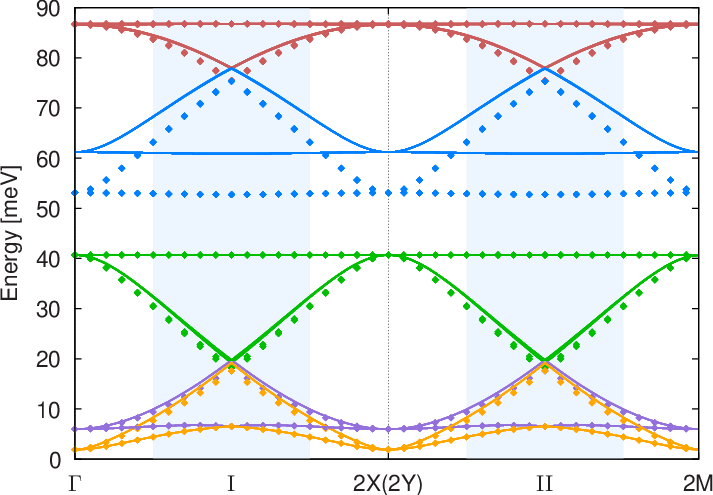}
 \caption{(color online). Bilayer: Effect of $\Fc$ on the quantum intermediate phase magnons. Non-zero magnetic couplings (in meV): $\Fp$(=$\F$)$=11.39$, $\A$($=\Ap$)$=1.5$, $\Az$($=\Bz$)$=0.88$, $\D=0.074$; while: $\Fc=0$ (symbols), and $\Fc=1.35$ (solid lines). $S_1=2$, $S_2=1.5$. $\theta = 0.$} 
 \label{FIG-12-n}
\end{figure}

\begin{figure}[H]
 \centering
 \includegraphics[width=0.6\columnwidth]{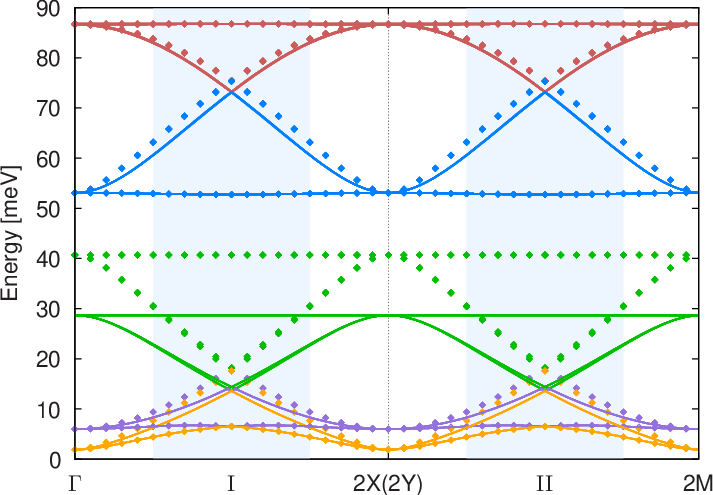}
 \caption{(color online). Bilayer: Effect of $\Ad$ on quantum intermediate phase magnons. Non-zero magnetic couplings, in meV: $\Fp$(=$\F$)$=11.39$, $\A(=$\Ap$)=1.5$, $\Az$($=\Bz$)$=0.88$, $\D=0.074$; while: $\Ad=0$ (symbols); and $\Ad=1.5$ (solid lines). $S_1=2$, $S_2=1.5$. $\theta = 0.$} 
 \label{FIG-13-n}
\end{figure}

To fit their INS data in bilayer \BCom{Pr}{Ca}{0.9}{Sr}{0.1}, Johnstone~\emph{et al.}~\cite{Johnstone} considered a localized spin model where they also included NNN couplings along the zig-zag chains: $\Fc$ and $\Ad$ in our notation of figure~\ref{FIG-1}. Interestingly, they mentioned that one good fit can be achieved if no charge disproportionation is asssumed ($S_1 = S_2$) and both $\Ad$ and $\Fc$ of almost identical magnitude are included: though with opposite signs, one of them AF and the other FM~\cite{Johnstone}. In this regard, they indicate that the excitation spectrum is unchanged by interchanging the values (and signs) of these two magnetic exchange parameters. In the Discussion section we address these points in detail. Here, as a complementary first step for that analysis, we illustrate the separate effects of these coupling parameters on the quantum intermediate phase magnons of bilayer compounds: in figure~\ref{FIG-12-n} we exhibit the effect of $\Fc$, and in figure~\ref{FIG-13-n} we show the effect of $\Ad$. Basically, we find that different magnon branches are affected by each of these two coupling parameters.

Concretely, figure~\ref{FIG-12-n} shows that increasing a ferromagnetic $\Fc$ mainly affects the magnon branches defining the top of the magnon gap (lifting it about 10 meV, in the case shown), while in the lower bands the effects are almost imperceptible. Furthermore, changing the sign of coupling $\Fc$ leads to the same magnon branches being affected, but the ensuing shift of energies has the opposite sign (and equal magnitude). On the other hand, figure~\ref{FIG-13-n} illustrates that increasing an antiferromagnetic $\Ad$ coupling, mainly shifts the magnon branches defining the bottom of the magnon gap, reducing their energy (by about 12 meV, in the case shown), while all other magnon branches are almost unchanged. Again, changing the sign of coupling $\Ad$ leads to the same magnon branches being affected, but the ensuing shift of energies has opposite sign (and equal magnitude).

Thus, our analysis demonstrates that the two coupling parameters $\Fc$ and $\Ad$ allow to fit the magnon gap, by tuning their magnitudes and signs. In particular, we have checked that in the absence of disproportionation, no magnon gap will exist if $\Fc = 0 = \Ad$. Furthermore, the separate effect of $\Fc$ and $\Ad$ on particular magnon branches up to now presented, allows to understand the statement of Johnstone~\emph{et al.}~\cite{Johnstone} regarding the possibility of interchanging their values and signs simultaneously for their fit. However, as discussed in more detail in section~\ref{sec.Discussion}, we found that a complex interplay of these two couplings with the antiferromagnetic inter-chain coupling parameter $\A$ exists in the model, so that in a more general case, with larger values of $\A$, one can not affirm that an interchange of $\Fc$ and $\Ad$ is always possible without affecting the magnon energies. 

\section{Effect of interchange of $\Fc$ and $\Ad$ couplings in \BCom{Pr}{Ca}{0.9}{Sr}{0.1}.}
\begin{figure}[H]
 \centering
  \includegraphics[width=0.6\columnwidth]{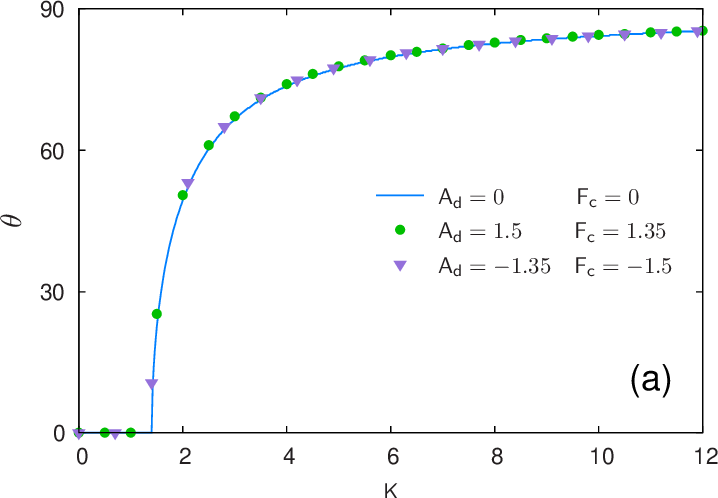}
  \\[5mm]
  \includegraphics[width=0.6\columnwidth]{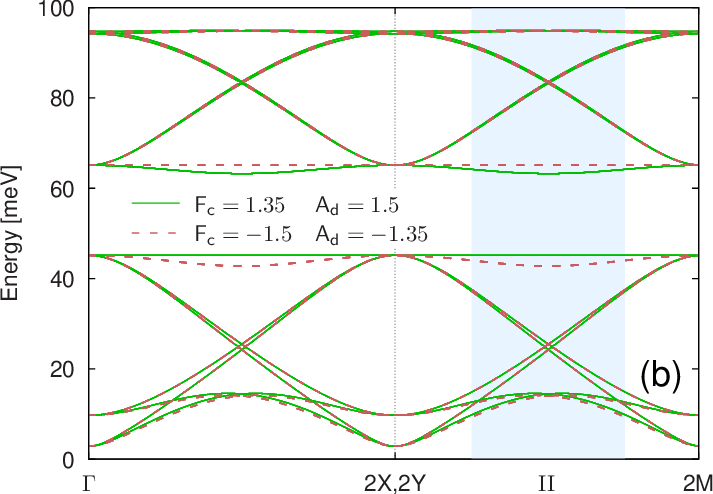}
  \caption{(color online) Bilayer: Effect of the NNN intra-chain couplings $\Fc$ and $\Ad$ with $S_1=S_2=1.75$. Non-zero magnetic couplings (in meV): $\Fp=11.39$, $\A=4.0$, $\Az$($=\Bz$)$=0.88$, $\D=0.074$, $\F$ and $\Ap$ as in equations~\eqref{EQ1.a}1.  $\Ad$ and $\Fc$ as detailed in inset. (a) Angle characteristic of the classical intermediate phase with minimal energy, as a function of $\K$. For: $\Ad=0$, $\Fc=0$ (solid line); $\Ad=1.5$, $\Fc=1.35$ (dots); $\Ad=-1.35$, $\Fc=-1.5$ (triangles). (b)~Magnons calculated with $\K=0, \theta=0$. For: $\Ad=1.5$, $\Fc=1.35$ (solid line); $\Ad=-1.35$, $\Fc=-1.5$ (dashed lines).} 
 \label{FIGS-14-n}
\end{figure}

Next, for bilayer compounds, we will discuss in detail the joint effect of the NNN intra-chain couplings $\Fc$ and $\Ad$ \. Especially, we analyze under which conditions it is possible to interchange their magnitudes and signs, as mentioned by Johnstone~\emph{et al.}~\cite{Johnstone}, without affecting the magnon spectrum. As mentioned in~\ref{sec.FcandAd} and seen in figure~\ref{FIGS-14-n}a, the angle $\theta$ that minimizes the classical energy and $\Kc$ do not change when we interchange the values and signs of $\Ad$ and $\Fc$, if the spin magnitudes are equal   ($S_1=S_2$), with parameters like Johnstone~\emph{et al.} had mentioned. However, we have found that the equivalence between both set parameters must be revised in the quantum case in general, and especially when $S_1\ne S_2$.

For instance, in figure~\ref{FIGS-14-n}b we compare the quantum magnons under the interchange of $\Fc$ and $\Ad$, for two cases with equal classical results (corresponding to $\K=0$ of  figure~\ref{FIGS-14-n}a\,) Notice that in Figs.\ref{FIGS-14-n} we have used $\A$\, larger than in the previous figures, in order to make the effect more visible. Here $S_1=S_2$, and even though the magnon spectra are quite similar, there are differences in some regions of the BZ (see shaded region, e.g.), and also the non-dispersive modes (flat magnon bands) of each case are located in different bands. Specifically, when $\Fc=1.35$ meV and $\Ad$=1.5 meV the flat band lies at the bottom of the magnon gap ($\sim$ 45 meV), while when we interchange the values and signs of these couplings, it is now placed at the top of the gap ($\sim$ 65 meV). It is interesting to stress that the differences found, allow to determine the appropriate signs for $\Ad$ and $\Fc$ in our model which, according to the experimental data~\cite{Johnstone}, would then correspond to an antiferromagnetic $\Ad $ and a ferromagnetic $\Fc$.

\begin{figure}[H]
  \centering
  \includegraphics[width=0.6\columnwidth]{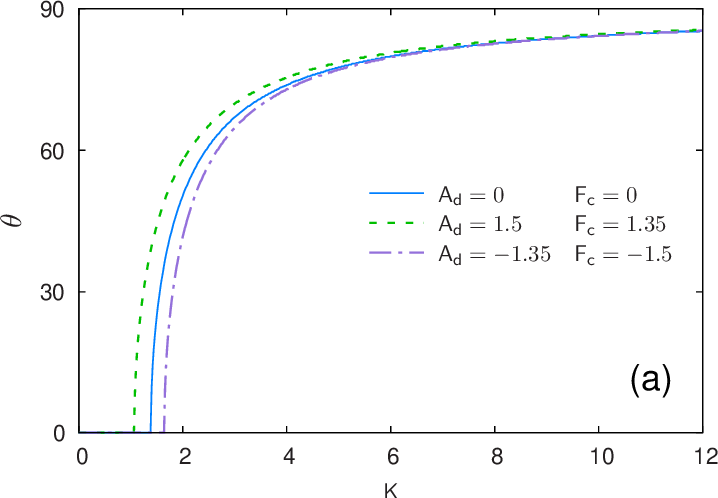}
  \\[5mm]
  \includegraphics[width=0.6\columnwidth]{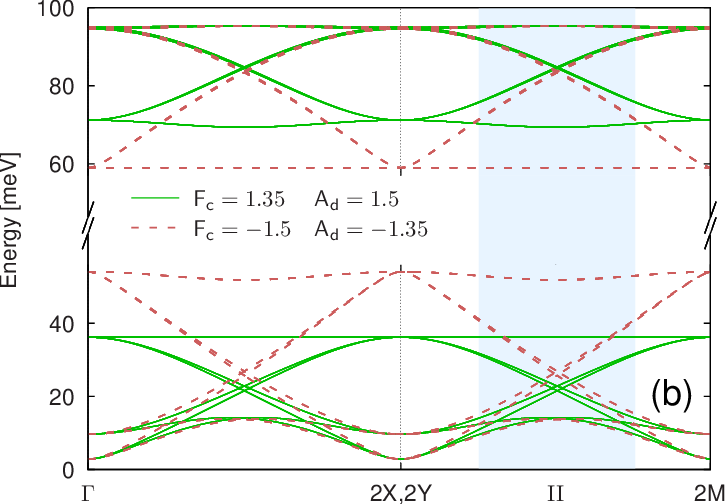}
  \caption{(color online). Bilayer: Effect of the NNN intra-chain couplings $\Fc$ and $\Ad$ with $S_1\neq S_2$ as in equations~\eqref{EQ1.b}. Non-zero magnetic couplings (in meV): $\Fp=11.39$, $\A=4.0$, $\Az$($=\Bz$)$=0.88$, $\D=0.074$, $\F$ and $\Ap$ as in equations~\eqref{EQ1.a}, $\Ad$ and $\Fc$ as detailed in inset. (a) Angle characteristic of the classically stable intermediate phase as a function of $\K$. For: $\Ad=0$, $\Fc=0$ (solid line), $\Ad=1.5$, $\Fc=1.35$ (dashed-line); $\Ad=-1.35$, $\Fc=-1.5$ (dash-dotted line). (b)~Magnons calculated with $\K=0$, $\theta=0$. For: $\Ad=1.5$, $\Fc=1.35$ (solid lines); $\Ad=-1.35$, $\Fc=-1.5$ (dashed lines).} 
  \label{FIGS-15-n}
\end{figure}

The effects discussed above, become amplified if the spins are of different magnitudes ($S_1\ne S_2$) and again using a larger value of $\A$, since this causes more dispersion in the bands near the top or the bottom of the gap. Figure~\ref{FIGS-15-n}a shows that with $S_1\ne S_2$, classically, angle $\theta$ and also $\Kc$ are different under interchange of $\Fc$ and $\Ad$. Not surprisingly, the corresponding quantum excitation spectra, which we show in figure~\ref{FIGS-15-n}b, differ even more between them: e.g. the shift observed in the flat band modes has been noticeably increased. 

\section*{References}
\bibliography{biblio}

\end{document}